\documentstyle[twoside,fleqn,col2,epsfig]{article}

\newcommand{\Ss}{\scriptstyle}
\newcommand{\Ts}{\textstyle}
\newcommand{\be}{\begin{eqnarray}}
\newcommand{\ee}{\end{eqnarray}}

\newcommand{\AmS}{{\protect\the\textfont2
  A\kern-.1667em\lower.5ex\hbox{M}\kern-.125emS}}

\title{Hyperons as collective excitations of 
chiral solitons\thanks{Talk presented at the III Int. Conf. on Hyperons, 
Charm and Beauty Hadrons, Genova, June--July 1998.}}

\author{Herbert Weigel\address{Institute for Theoretical Physics,
        T\"ubingen University\\
        Auf der Morgenstelle 14, D--72076 T\"ubingen, Germany}%
        \thanks{Heisenberg--Fellow}%
        \thanks{Address after Sept. $1^{\rm st}$ 1998: 
         Center for Theoretical Physics,
         Laboratory for Nuclear Science and Dept. of Physics,
         Massachusetts Institute of Technology,
         Cambridge, Mass. 02139.}
         }
       
\begin{document}

\begin{abstract}
According to the large $N_C$ limit of QCD baryons are considered
as soliton solutions in effective mesons theories. While the 
classical solitons dwell in the isospin subgroup of flavor SU(3) 
hyperon states are generated by canonical quantization of the 
collective coordinates which describe the flavor orientation
of the soliton. The resulting Hamiltonian is diagonalized 
exactly allowing one to discuss the dependence of various 
baryon properties on flavor symmetry breaking. In particular 
axial charges, baryon magnetic moments and radiative decay widths 
are considered.

\end{abstract}

\maketitle

\section{INTRODUCTION}

The generalization of QCD to an arbitrary large number ($N_C$) of color 
degrees of freedom indicates that for $N_C\to\infty$ QCD becomes 
equivalent to a non--linear effective theory ${\cal A}$ of weakly 
interacting mesons, as the associated meson coupling constants vanish 
in this limit. The action ${\cal A}$ may contain soliton solutions and 
the dependences of their properties on $N_C$, inherited from the meson 
coupling constants, is that of baryons in large $N_C$ QCD. Hence the soliton 
solutions are identified as baryons \cite{Wi79}. The building block to 
construct ${\cal A}$ for low energies is the concept of chiral symmetry 
and its spontaneous breaking. Therefore the chiral field, {\it i.e.} 
the non--linear realization of the pseudoscalar fields 
($\Phi^a=\pi,k,\eta$) 
\be
U={\rm exp}\left(i\lambda^a\Phi^a/f_a\right), 
\label{eq_1}
\ee
is the relevant degree of freedom. Here $\lambda^a$ are the SU(3) 
Gell--Mann matrices and $f^a$ denote pertinent decay constants.
The most prominent chirally invariant model containing solitons 
is the so--called Skyrme model \cite{Sk61}. This model can be 
generalized to the three flavor case \cite{Gu84} with the explicit 
breaking flavor symmetry included (for a review see \cite{We96}). 
The essential ingredient is that the strange current quark mass is 
smaller than $\Lambda_{\rm QCD}$. 

To minimize the soliton mass, $M_{\rm cl}$ in the three flavor 
model the static hedgehog must be embedded in the isospin subspace:
\vspace{-0.1cm}
\be
U_H({\vec r})={\rm exp}
\left(i\sum_{i=1}^3 {\hat r}_i\lambda^i F(r)\right).
\label{eq_2}
\ee
Subsequently the zero--modes of $U_H$ are assumed to be time dependent:
\be 
U({\vec r},t)= A(t)\, U_H({\vec r})\, A^\dagger (t).
\label{eq_3}
\ee 
The quantum mechanical treatment of the collective coordinates 
$A(t)\in {\rm SU(3)}$ leads to states which are identified as the 
low--lying $\frac{1}{2}^+$ and $\frac{3}{2}^+$ baryons. This treatment 
is called the rigid rotator approach (RRA) to include strangeness.

\section{BARYON STATES WITH FLAVOR SYMMETRY BREAKING}

Substituting the {\it ansatz} (\ref{eq_3}) yields the
Lagrangian for the collective coordinates $A$
\vspace{-0.1cm}
\be
L(A,\dot A)&=&-M_{\rm cl}+\frac{1}{2}\alpha^2\sum_{i=1}^3\Omega_i^2
+\frac{1}{2}\beta^2\sum_{\alpha=4}^7\Omega_\alpha^2
\nonumber \\ &&
-\frac{\sqrt3}{2}\Omega_8
-\frac{1}{2}\gamma\left(1-D_{88}\right)+\ldots\, .
\label{eq_4}
\ee
The angular velocities $\Omega_a$ and the adjoint representation $D_{ab}$ 
are defined as
\be
A^\dagger \dot A = \frac{i}{2}\sum_{a=1}^8 \lambda^a\Omega_a
\, ,\, \,
D_{ab}=\frac{1}{2}\, 
{\rm tr}\left(\lambda_a A\lambda_b A^\dagger\right) .
\label{eq_5}
\ee
The coefficients $\alpha^2,\beta^2$ and $\gamma$ are functionals 
of the soliton. Their actual values depend on the details 
of the considered meson theory \cite{We96}. The contribution linear
in $\Omega_8$ stems from the Wess--Zumino action which mocks up
the axial anomaly in the effective meson theory \cite{Wi83}. The term 
including the SU(3)--D--function is due to flavor symmetry 
breaking. The ellipsis in eq (\ref{eq_4}) refer to symmetry breaking
terms which are subject to the specific model.

\begin{figure}[htb]
\hspace{-0.7cm}
\epsfig{file=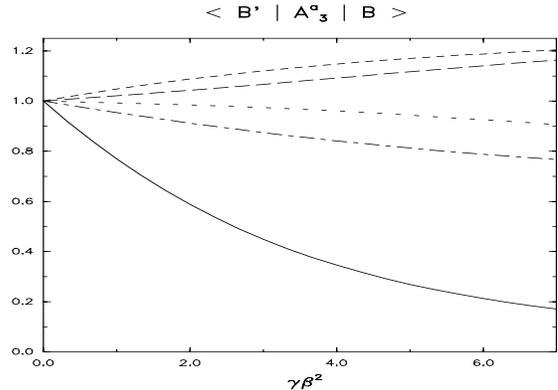,height=5.0cm,width=7.8cm}
\vspace{-0.9cm}
\caption{Axial current matrix elements as a function of 
the effective symmetry breaking parameter
${\Ss \gamma\beta^2}$. Full line:
${\Ss \langle p|{\overline{s}}\gamma_3\gamma_5s| p\rangle}$;
dashed dotted line:
${\Ss \langle p|{\overline{u}}\gamma_3\gamma_5s| \Lambda\rangle}$;
dotted line:
${\Ss \langle n|{\overline{u}}\gamma_3\gamma_5s|
\Sigma^-\rangle}$;
long dashed line:
${\Ss \langle\Lambda|{\overline{u}}\gamma_3\gamma_5s|
\Xi^-\rangle}$;
dashed line:
${\Ss \langle p|{\overline{u}}\gamma_3\gamma_5d|
n\rangle}$; These matrix elements, which are taken from refs
\protect\cite{Pa89a} and \protect\cite{Pa90}, are normalized to the
flavor symmetric values.}
\label{fig_1}
\vspace{-0.7cm}
\end{figure}

When quantizing the collective coordinates one identifies the right 
generators of SU(3):
$R_a=-\partial L/\partial \Omega_a$. The first three generators
actually are (up to a sign) the spin--operators. The constraint 
$R_8=\sqrt3 / 2$ requires half--integer spin eigenstates. 
The Hamiltonian  $H=-\sum_{a=1}^8R_a\Omega_a -L =H_0+H_{\rm SB}$ 
is conveniently separated into flavor symmetric ($H_0$) and 
symmetry breaking ($H_{\rm SB}$) pieces. 
For $H_{\rm SB}=0$ the eigenstates are members 
of a certain SU(3) representation, {\it e.g.} the octet 
for the $\frac{1}{2}^+$ baryons. Due to flavor symmetry 
breaking these states acquire contributions from higher 
dimensional representations \cite{Pa89}, {\it e.g.}
\be
|N\rangle&=&|N,{\bf 8}\rangle 
+ 0.075 \gamma\beta^2 |N,{\overline{\bf 10}}\rangle
\nonumber \\ &&\hspace{1cm}
+ 0.049 \gamma\beta^2 |N,{\bf 27}\rangle+\ldots
\\
|\Lambda\rangle&=&|\Lambda,{\bf 8}\rangle
+ 0.060 \gamma\beta^2 |\Lambda,{\bf 27}\rangle+\ldots
\label{eq_6}
\ee
for the nucleon and the $\Lambda$ hyperon. The higher order perturbation 
pieces have been indicated. Similarly the shift of the baryon mass due 
to symmetry breaking is computed. For the nucleon one finds
\be
\vspace{0.1cm}
\delta M_N=\frac{-1}{2\beta^2}\left[0.3\gamma\beta^2
+0.029\left(\gamma\beta^2\right)^2+\ldots\right]\, .
\label{eq_7}
\vspace{0.1cm}
\ee
Obviously the product $\gamma\beta^2$ is the effective symmetry 
breaker rather than only $\gamma$. Typical values are 
$\gamma\beta^2\approx 3.0 - 4.0$.
Actually the collective Hamiltonian can be diagonalized exactly 
by means of an ``Euler angle'' representation for the collective 
coordinates $A$ \cite{Ya88}. A typical result for the baryon spectrum 
is displayed in table \ref{tab_1}.
\begin{table*}[hbt]
\setlength{\tabcolsep}{1.5pc}
\newlength{\digitwidth} \settowidth{\digitwidth}{\rm 0}
\catcode`?=\active \def?{\kern\digitwidth}
\caption{Baryon mass differences with respect to the nucleon. Data
are in MeV. Results are from \protect\cite{We96}.}
\label{tab_1}
\begin{tabular*}{\textwidth}{@{}l|@{\extracolsep{\fill}}ccccccc}
\hline
& $\Ts \Lambda$ & $\Ts \Sigma $ & $\Ts \Xi$ & $\Ts \Delta $
& $\Ts \Sigma^* $ & $\Ts \Xi^* $ & $\Ts \Omega $ \\
\hline
Model & 163 & 264 & 388 & 268 & 410 & 545 & 680 \\
Expt. & 177 & 254 & 379 & 293 & 446 & 591 & 733 \\
\hline
\end{tabular*}
\vspace{-0.2cm}
\end{table*}
The mass differences are reasonably well reproduced, on the 10\% level. 
Many static properties can be computed once the 
baryon states have been constructed from $H$. In particular 
the variation with flavor symmetry breaking can be studied.
An example is provided in figure \ref{fig_1} where various 
axial current matrix elements are displayed.
Although the flavor changing axial current matrix elements,
which enter the Cabibbo model for hyperon beta decay, vary only 
moderately with symmetry breaking the diagonal matrix element 
$\langle p|{\overline{s}}\gamma_3\gamma_5s| p\rangle$ gets reduced 
to approximately half its flavor symmetric value. Hence SU(3) is 
eventually a good symmetry to relate various beta decay matrix 
elements, however, it seems dangerous to assume SU(3) symmetry 
to extract $\langle p|{\overline{s}}\gamma_3\gamma_5s| p\rangle$
from data on hyperon beta decay. This has frequently been done 
in the context of the {\it proton spin puzzle} and yielded an
unexpectedly large amount of polarized strange quarks in the 
nucleon.

\section{THE SLOW ROTATOR}

Although the exact diagonalization of the collective Hamiltonian
includes the effects of flavor symmetry breaking not all observed
deviations from the symmetric formulation are reproduced. An example
is given in table \ref{tab_2} where ratios of magnetic moments are 
compared to experiment. Clearly, the RRA underestimates the observed 
deviation from unity, which is predicted in the symmetric formulation.
This is linked to the inability of the RRA to include the influence
of symmetry breaking on the size of the classical soliton. This
short--coming is cured within the slow rotator approach (SRA)
\cite{Sch91}. Making explicit the strangeness changing angle $\nu$
in an ``Euler angle'' parameterization of $A$ ($R_2$ are 
SU(2) matrices):
\vspace{-0.1cm}
\be
A=R_2(\alpha,\beta,\gamma)\, {\rm e}^{-i\nu\lambda^4}
R_2(\alpha^\prime,\beta^\prime,\gamma^\prime)\, 
{\rm e}^{-i\rho\lambda^8/\sqrt{3}}
\hspace{-2pt}
\label{eq_8}
\ee
yields the (static) Lagrangian 
\vspace{-0.1cm}
\be
L_\nu[F]=-M_{\rm cl}[F]-\frac{3}{4}\gamma[F]\, {\rm sin}^2 \nu \, .
\label{eq_9}
\ee
Minimizing $L_\nu[F]$ for a given $\nu$ causes the chiral angle $F$ to 
parametrically depend on the strangeness orientation: It decays with 
the pion mass for $\nu=0$ while the configuration which is maximally 
rotated into strange direction ($\nu=\pi/2$) has the kaon mass 
entering the Yukawa tail. This additional dependence on the 
collective coordinates must be taken into account when computing 
static baryon properties \cite{Sch91}. As a result the desired
deviation from the symmetric results in the
magnetic moments in achieved as shown in table \ref{tab_2}.
\vspace{-0.6cm}
\begin{table}[hbt]
\setlength{\tabcolsep}{1.5pc}
\catcode`?=\active \def?{\kern\digitwidth}
\caption{U--spin relations for magnetic moments of
$\Ts \frac{1}{2}^+$ baryons.}
\label{tab_2}
\begin{tabular}{@{}l|@{\extracolsep{1pc}}c
@{\extracolsep{1pc}}c
@{\extracolsep{1pc}}c}
\hline
&$\Ts \mu_{\Sigma^+}/\mu_p$ & $\Ts \mu_{\Xi^0}/\mu_n $ &
$\Ts \mu_{\Xi^-}/\mu_{\Sigma^-} $ \\
\hline
RRA   & 0.98 & 0.97 &  0.79 \\
Expt. & 0.87 & 0.66 &  0.59 \\
SRA   & 0.85 & 0.65 &  0.50 \\
\hline
\end{tabular}
\end{table}
\vspace{-1.4cm}
\begin{table}[hb]
\setlength{\tabcolsep}{1.5pc}
\catcode`?=\active \def?{\kern\digitwidth}
\caption{Widths for radiative hyperon decays. Results are 
normalized to the width of $\Ts \Delta\to\gamma N$.}
\label{tab_3}
\begin{tabular}{@{}l|@{\extracolsep{1pc}}c
@{\extracolsep{1pc}}c
@{\extracolsep{1pc}}c}
\hline
& SRA & RRA & SU(3)--sym. \\
\hline
$\Ts \Sigma_0^*\to\gamma\Lambda $ \hspace{-0.2cm}
& 0.509 & 0.653 & 0.75 \\
$\Ts \Sigma_-^*\to\gamma\Sigma_-$ \hspace{-0.2cm}
& 0.005 & 0.007 &   0  \\
$\Ts \Sigma_0^*\to\gamma\Sigma_0$ \hspace{-0.2cm}
& 0.024 & 0.035 & 0.25 \\
$\Ts \Sigma_+^*\to\gamma\Sigma_+$ \hspace{-0.2cm}
& 0.152 & 0.210 &  1   \\
$\Ts \Xi_-^*\to\gamma\Xi_-$       \hspace{-0.2cm}
& 0.004 & 0.011 &  0   \\
$\Ts \Xi_0^*\to\gamma\Xi_0$       \hspace{-0.2cm}
& 0.204 & 0.313 &  1   \\
\hline
\end{tabular}
\vspace{-0.5cm}
\end{table}

\noindent
The predicted baryon radii decrease by about 15\% per unit strangeness
in agreement with empirical observations \cite{Po90}. In the SRA the
widths for the radiative decays of the $\frac{3}{2}^+$ baryons have
been calculated \cite{Ha97}. In table \ref{tab_3} the results are
compared to the RRA and SU(3) symmetric predictions.
Again, sizeable reductions with decreasing strangeness are observed.
However, the U--spin predictions \cite{Li92}
$\Gamma(\Sigma_-^*\to\gamma\Sigma_-)=0$ and
$\Gamma(\Xi_-^*\to\gamma\Xi_-)=0$ are maintained.
A similarly strong dependence on strangeness is found for the
electric polarizabilities.
The predictions in the magnetic case
are small due to the cancellation between dispersive and
seagull pieces \cite{Sco96}.

\section{CONCLUSIONS}

Considering hyperons as collective excitations of chiral 
solitons has provided an effective means to study the 
influence of flavor symmetry breaking on various baryon
matrix elements. In particular I have shown that relating
$\langle p|{\overline{s}}\gamma_3\gamma_5s| p\rangle$ to 
data on hyperon beta decay is suspicious, a result 
also obtained in other models \cite{Li95}. The pattern
of the baryon magnetic moments requires one to include the 
effect of symmetry breaking in the determination of the soliton 
profile. This generally yields a sizable decrease of 
flavor conserving baryon matrix elements with strangeness.

\end{document}